\documentclass[11pt]{article}

\usepackage{color}
\usepackage{fancyvrb}

\DefineVerbatimEnvironment{Highlighting}{Verbatim}{commandchars=\\\{\}}
\usepackage{framed}
\definecolor{shadecolor}{RGB}{241,243,245}

\usepackage{xcolor} 
\usepackage[numbers]{natbib}
\usepackage{graphicx}
\usepackage{amsmath, amsfonts, amssymb, mathrsfs, rotating, setspace}
\usepackage[utf8]{inputenc}
\usepackage[margin=2.5cm]{geometry}
\usepackage{url} 
\usepackage{tikz}
\usepackage{rotating}
\usepackage{adjustbox}
\usepackage{longtable}
\usepackage{booktabs}

\usetikzlibrary{calc}
\usetikzlibrary{positioning}
\usetikzlibrary{arrows}

\newcommand\independent{\protect\mathpalette{\protect\independenT}{\perp}}
\def\independenT#1#2{\mathrel{\rlap{$#1#2$}\mkern2mu{#1#2}}}

\title{Inverse probability of treatment weighting with generalized linear outcome models for doubly robust estimation}

\author{Erin E Gabriel* \and 
Michael C Sachs \and 
Torben Martinussen \and 
Ingeborg Waernbaum \and 
Els Goetghebeur \and 
Stijn Vansteelandt \and 
Arvid Sjölander} 

\begin{document}

\maketitle

\abstract{
There are now many options for doubly robust estimation; however, there is a concerning trend in the applied literature to believe that the combination of a propensity score and an adjusted outcome model automatically results in a doubly robust estimator and/or to misuse more complex established doubly robust estimators. A simple alternative, canonical link generalized linear models (GLM) fit via inverse probability of treatment (propensity score) weighted maximum likelihood estimation followed by standardization (the g-formula) for the average causal effect, is a doubly robust estimation method. Our aim is for the reader not just to be able to use this method, which we refer to as IPTW GLM, for doubly robust estimation, but to fully understand why it has the doubly robust property. For this reason, we define clearly, and in multiple ways, all concepts needed to understand the method and why it is doubly robust. In addition, we want to make very clear that the mere combination of propensity score weighting and an adjusted outcome model does not generally result in a doubly robust estimator. Finally, we hope to dispel the misconception that one can adjust for residual confounding remaining after propensity score weighting by adjusting in the outcome model for what remains `unbalanced' even when using doubly robust estimators. We provide R code for our simulations and real open-source data examples that can be followed step-by-step to use and hopefully understand the IPTW GLM method. We also compare to a much better-known but still simple doubly robust estimator. \\
\textit{keywords:} causal inference, doubly robust, generalized linear models}


\clearpage
\section{Introduction} 
Recently it has become well-known that many commonly used outcome models have the property that, within clinical trials where there is assumed to be no confounding, maximum likelihood in combination with regression standardization or the g-formula delivers unbiased estimates of the average causal effect (ACE) of the intervention (in large samples) even if the model is misspecified. Specifically, this is true for canonical link generalized linear models (GLM) \citep{rosenblum2009using, tackney2023comparison}. 
  
Standardization estimators based on canonical GLMs fitted using weighted maximum likelihood estimation (MLE) with propensity score weights were shown to be doubly robust for the ACE in observational settings\cite{robinscomment2007} when there are no unmeasured confounders. Throughout, we will refer to these estimators as IPTW GLM or IPTW OLS, although we note that the model itself is not being weighted, but rather the equations used to estimate the model parameters, and that we further regression standardize after fitting the weighted MLE. Regression standardization, also called regression imputation and g-computation \cite{ROBINS19861393, keil2014parametric}, is the averaging over the observed or known distribution of the covariates included in a regression model while holding the exposure constant. Robins et al.\cite{robinscomment2007}, in a commentary on Kang et al.\cite{kang2007demystifying}, credits the first recognition of the double robustness property of IPTW GLM to Marshall Joffe, who never published these results. To our knowledge, IPTW GLM or IPTW OLS is not often used in the applied literature in observational settings and not well known in general. Another doubly robust estimator that was made accessible by Funk et al.\cite{funk2011doubly}, but developed by Robins et al.\cite{robins1994estimation}, is much more widely used. It uses an unweighted outcome model, the predictions from which are combined with the propensity score weights in a weighted average. 

The reason for the difference in use between IPTW GLM and the method described in Funk et al.\cite{funk2011doubly}, may be due to the former not being widely known outside of the field of causal inference or the lack of examples of easy-to-use code for applying it. It is also possible that worries about highly biased estimation when both the models for the propensity and the outcome are misspecified may prevent applied researchers from using IPTW GLM. However, this final reservation would equally apply to the Funk et al.\cite{funk2011doubly} method and to IPTW GLM. 

Although these methods are not new, they appear to be largely unknown in the applied community. Also, to our knowledge, these methods have only been discussed in theoretical journals and contexts (e.g., the commentary by Robins et al.\cite{robinscomment2007}). Thus, in this paper we attempt to provide an accessible introduction to these methods. We also focus on the limitations of these methods, e.g., by clearly defining the scenarios where they do not provide double robustness. Specifically, we aim to clearly define all terms needed to understand why IPTW canonical link GLMs fit via maximum likelihood or IPTW OLS result in DR estimators, and why in contrast the resulting estimators from IPTW non-canonical link GLM are not, in general, doubly robust. We offer alternatives to the IPTW non-canonical link GLM estimators that target the same parameters of interest. We show that IPTW GLM and the estimator suggested in Funk et al.\cite{funk2011doubly} are asymptotically equivalent when both models are correctly specified. We also show that both methods are asymptotically efficient when both models are correctly specified, as they are (up to asymptotic equivalence) an estimator suggested by the efficient influence function, as they render the sample average of the efficient influence functions (evaluated at the estimators) zero.We compare these two estimators in finite sample simulations and find very similar estimates and efficiency in simulations. We provide real data analyses to demonstrate how to use the IPTW GLM and IPTW OLS. Finally, we provide example R code to implement the IPTW canonical link GLM followed by regression standardization as well as R code for the DR estimator suggested in Funk et al.\cite{funk2011doubly}.

\section{Notation and Preliminaries} \label{not}
 Let $Y$ be the outcome of interest, and let $X$ be the exposure of interest, which we will assume to be binary throughout. Let $Y(x)$ be the counterfactual or potential value of $Y$ for a particular subject had the exposure, potentially counter to fact, been set to $x$. 
Let $\boldsymbol{Z}=\{Z_1, \ldots, Z_m\}$ be a set of measured confounders, conditional on which there is no residual confounding. Thus, all confounders are measured sufficiently to allow for inclusion in an outcome or propensity score model using some functional form only depending on observed data. Let the target of causal inference be the ACE which is defined in counterfactual terms as $E\{Y(1)\}-E\{Y(0)\}$.  

\subsection{Terms and concepts}
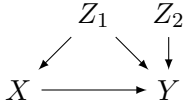
\begin{figure*}[ht]
\centering
\begin{tikzpicture}
\node (v) at (0,0) {$X$};
\node (i) at (2,0) {$Y$};
\node (uh) at  (1,1) {$Z_1$};
\node (uh2) at (2,1) {$Z_2$};
\draw[-latex] (uh) -- (i);
\draw[-latex] (uh2) -- (i);
\draw[-latex] (uh) -- (v);
\draw[-latex] (v) -- (i);
\end{tikzpicture}
\caption{Causal diagram for an exposure $X$, outcome $Y$, confounder $Z_1$, and a precision variable or interaction variable $Z_2$ that only affects $Y$.  \label{DAGS}}
\end{figure*}

In non-randomized settings, there is a concern that we are finding an association between the outcome and the exposure because there are common causes that lead to both the outcome and the exposure rather than a causal effect of the exposure on the outcome. This is depicted in the causal diagram in Figure \ref{DAGS}. If one is only specifying a model for the outcome, then to estimate the causal effect of interest, one must, in general, correctly specify the relationships between the outcome and a sufficient set of confounders to control confounding; please see Witte et al.\cite{witte2019covariate} for more detail on what constitutes a sufficient adjustment set. For example, in the setting in Figure \ref{DAGS}, $Z_1$ is a confounder, but $Z_2$ is not. One would need to correctly account for the relationship between $\{Z_1,X\}$ and $Y$ to estimate the causal effect of $X$ on $Y$ using an outcome model, but the inclusion of $Z_2$ is not needed.  We will assume that there will always be a correct, but potentially very complex, specification of some model that contains the exposure of interest and a sufficient set of confounders; it just may not be of the model type one is considering or even a standard regression model.

We will use the terms `misspecified model', `misspecification', and `correct model' throughout. We will be specific here to avoid any confusion. Let the distribution of the true data-generating mechanism of $Y$ given $X$ and $\boldsymbol{Z}$ be $f(y;X,\boldsymbol{Z})$, then a regression model for $Y$, say $s(X,\boldsymbol{Z};\beta)$, which corresponds to some set of probability distributions, $\Omega$, is correctly specified if it contains the true distribution, i.e.
\begin{eqnarray}
\label{eq:correct}
 f(y; X,\boldsymbol{Z}) \in \Omega.
\end{eqnarray}
For example, let $s(X,Z;\beta)$ be a model for the mean, and given the true distribution $f(y;X,Z)$, let the true conditional mean be $E(Y|X,Z)$. Then $s(X,Z;\beta)$ is correctly specified if for some $\beta^*$, $s(X,Z, \beta^*)=E(Y|X,Z)\textrm{ for all } (X,Z).$ This can been seen easily in a simple all binary example. For a binary $Y$, $X$, $Z_1$ and $Z_2$, the mean model for $Y$ containing any subset of $\{X,Z_1,Z_2\}$, and all interactions within that subset, will be statistically correctly specified because it will be saturated. Thus, multiple models may be simultaneously correctly specified. We note that although our example only considers the specification of the mean, if one makes assumptions about the error distribution, such as using the model-based standard error estimates, this can also be misspecified. In what follows we will only consider the correct specification of the mean model.

There are two things of note. First, a model need not depend on $\boldsymbol{Z}$ to be correctly specified. A model $s(X)$ can be correctly specified if it does not conflict with true data generating mechanism $f(X,\boldsymbol{Z})$, i.e., if $f(X,\boldsymbol{Z})$ is among the probability laws that are possible for $Y$ under the assumption that $s(X)$ is true. Second, this definition makes no reference to causality. Thus, the model may be correctly specified even if $\boldsymbol{Z}$ is not sufficient for confounding control. If $\boldsymbol{Z}$ is sufficient for confounding control, then we have that 
\begin{eqnarray}
\label{eq:suff}
Y(x) \independent X|\boldsymbol{Z},
\end{eqnarray}
so that $E(Y|X=x,\boldsymbol{Z})=E(Y(x)|\boldsymbol{Z})$. When both (\ref{eq:correct}) and (\ref{eq:suff}) hold we say that the model is correctly specified for confounding. Going back to the setting in Figure \ref{DAGS}, this would mean that the outcome model for $Y$ is correctly specified and is a nontrivial function of $Z_1$. This correct specification must contain and correctly specify the relationship between $Y$ and $\{X,Z_1\}$, to be correctly specified for confounding, but it may or may not need to include $Z_2$. We note that this definition implies that $Z$ contains no mediators, as we are discussing the correct specification for confounding specifically such that $E(Y|X=x,\boldsymbol{Z})=E(Y(x)|\boldsymbol{Z})$, and not for the counterfactual $Y(x,z)$. For greater discussion of this point and the ``doubly robust paradox" we refer readers to Keil et al. 2018\cite{Keilparadox}. Mediators are not the only variables that need to be considered when forming an adjustment set, as colliders should also not be included. For a detailed rigorous discussion of what constitutes a sufficient adjustment set we direct readers to \textit{Causal inference: What if}\cite{hernan2020causal}.

Correct specification and correct specification for confounding can occur in several ways, particularly if there are a large number of potential confounders where different sets would be sufficient for confounding control. This is particularly true for a collapsible model and estimand. For example, linear regression may be correctly specified simultaneously by including only one continuous variable in the correct functional form or the exposure and additional variables, also in the correct functional form, or any subset of these covariates, e.g. with squared or logged versions of the measured variables. In contrast, if a logistic regression model including two continuous variables is correctly specified, then it is highly unlikely that marginalizing out one of the variables also results in a logistic model \cite{gail84}. Odds ratios are also not collapsible, which adds additional complexity. 

An estimand, or parameter, obtained from a particular regression model being collapsible is another concept that has received much attention and been given multiple slightly different, or completely different, definitions in the literature, a typical  example of noncollapsibility is that of the odds ratio. Here, we will follow Neuhaus et al.\cite{neuhaus1993geometric}, and  Daniel et al.\cite{daniel2021making} and define collapsibility as follows. For a link function $q$, and a mean model $q\{E(Y|X,Z_1)\} = \beta_0 +\beta_1 X + \beta_2 Z_1$, the function that maps $E(Y|X=1,Z_1)$ to $E(Y|X=0,Z_1)$ is called by Daniel et al.\cite{daniel2021making} the ` characteristic collapsibility function'. This function is given by $h(v)=q^{-1}\{q(v) +\beta_1\}$, and regardless of the link function, $E(Y|X=1,Z_1)=h((E(Y|X=0,Z_1))$. If in addition, $$E\{h(E(Y|X=0,Z_1))\} = h(E\{E(Y|X=0,Z_1)\})$$ 
then the model/estimand combination is collapsible. Note that the definition as given requires that the conditional association between $X$ and $Y$, conditional on $Z_1=z$, does not depend on $z$, i.e. no interactions. It follows from this definition that (non-)collapsibility is a feature of the mean model's link function specifically.

The definition simplifies when the included covariate is independent of $X$, for example, in the setting in Figure \ref{DAGS} $Z_2\bot X$. Here, $Z_2$ may be a predictor for $Y$ but not a confounder for the $X$-$Y$ association. Assume further that $q\{E(Y|X,Z_2)\}=\beta_0+\beta_1 X+\beta_2 Z_2$ for some link function $q$, i.e., that the conditional association between $X$ and $Y$ given $Z_2$, on the scale defined by the link function $q$, is constant ($=\beta_1$) over levels of $Z_2$. We then say that the association parameter for $X-Y$, $\beta_1$, is collapsible if $q\{E(Y|X)\}=\beta^{'}_0+\beta_1 X$, i.e., if the marginal (over $Z_2$) association between $X$ and $Y$, on the scale defined by $q$, is equal to the constant conditional association $\beta_1$. This follows from our definition above 
because $E\{E\{Y|X=0,Z_2\}\}=q^{-1}(\beta'_0)$, and $h(q^{-1}(\beta'_0))=q^{-1}(\beta'_0+\beta_1)$, which is equal to $E\{h(E\{Y|X=0,Z_2\})\} = E\{E\{Y|X=1,Z_2\}\}=q^{-1}(\beta'_0+\beta_1)$.

As noted in the introduction, regression standardization\citep{tan2010bounded} is marginalizing over the observed or known distribution of the included covariates while holding the exposure constant. This allows for estimation of marginal causal effects from models with interactions between the exposure and the covariates, and from noncollapsible estimand/models including covariates. We note that the form of regression standardization we consider marginalizes over the observed covariate distribution, rather than some alternative distribution in order to transport the causal effect to a new population. For example, standardization can be used to obtain estimates of the marginal causal risk ratio using the predictions from a multivariable logistic regression in the current population. In what follows, all estimators we will consider involve regression standardization, although we note that when an estimand/model combination is collapsible, there is no need for regression standardization if there are no interactions with the exposure in the outcome model, as the coefficient for the exposure can also be interpreted as a marginal causal effect. Going back to the mean model above, $q\{E(Y|X,Z_1)\} = \beta_0 +\beta_1 X + \beta_2 Z_1$, and assume we fit this model via maximum likelihood to obtain $\hat{\beta}=(\hat{\beta}_0, \hat{\beta}_1, \hat{\beta}_2)$. The standardization estimator takes the form 
 \begin{eqnarray*}
\widehat{E\{Y(x)\}}=\hat{E}\{\hat{E}(Y|X=x,\boldsymbol{Z})\}=\frac{1}{n} \sum_{i=1}^n q^{-1}(\hat{\beta}_0 + \hat{\beta}_1X_i + \hat{\beta}_2Z_{1i})=\frac{1}{n} \sum_{i=1}^n s(x,\boldsymbol{Z}_i;\boldsymbol{\hat{\beta}})
\end{eqnarray*}
where for generality $s(\cdot)$ is an arbitrary but known and fixed parameterization of $E(Y|X=x,\boldsymbol{Z})$. The standardization estimator for the ACE is then given by: 
 \begin{eqnarray*}
\widehat{E\{Y(1)\}}-\widehat{E\{Y(0)\}}=\frac{1}{n} \left\{\sum_{i=1}^n s(1,\boldsymbol{Z}_i;\boldsymbol{\hat{\beta}}) -  \sum_{i=1}^n s(0,\boldsymbol{Z}_i;\boldsymbol{\hat{\beta}})\right\}.
\end{eqnarray*}

Finally, we will discuss generalized linear regression outcome models as those with and without a canonical link function. To avoid confusion about link functions, we outline the common canonical link functions in Table \ref{cl} below. A canonical link function transforms the mean of an outcome so that it can be represented by the natural exponential (location) parameter for the exponential family of distributions; we refer readers to \textit{Generalized Linear Models}\cite{mccullagh2019generalized} pages 26-32 for a detailed description. Further discussion of canonical links and what this means in the context of GLMs is also provided in the supporting information.

\begin{table}[h]
    \centering
    \begin{tabular}{c|c}
    Error Distribution & Canonical Link \\
    \hline 
      binomial& $q(\mu)=\log(\mu/(1-\mu))$\\
Gaussian&$q(\mu)=\mu$ \\
gamma&$q(\mu)=1/\mu$\\
inverse Gaussian&$q(\mu)=1/\mu^2$\\
Poisson&$q(\mu)=\log(\mu)$\\
    \end{tabular}
    \caption{Canonical link functions \label{cl}}
\end{table}

\section{IPTW Ordinary Least Squares}
Consider the working outcome model given by:
 \begin{eqnarray}
E(Y|X,\boldsymbol{Z})=\gamma_0+\beta X+m(\boldsymbol{Z};\boldsymbol{\gamma}), \label{linnoint}
\end{eqnarray}
where $m(\boldsymbol{Z};\boldsymbol{\gamma})$ is an arbitrary known function of $\boldsymbol{Z}$, and not $X$, for example $\gamma_1Z_1+ \gamma_2Z_2+\gamma_3Z_3$. We note that the term `working' here is used when proposing a model that is not necessarily assumed to be correctly specified, which is the standard for doubly robust estimators.

The propensity of treatment or non-treatment weights $$W(X,\boldsymbol{Z};\hat{\boldsymbol{\alpha}})=\frac{X}{g(\boldsymbol{Z};\hat{\boldsymbol{\alpha}})}+\frac{1-X}{1-g(\boldsymbol{Z};\hat{\boldsymbol{\alpha}})},$$
are based on the working model propensity model, 
\begin{eqnarray}
p(X=1|\boldsymbol{Z})=g(\boldsymbol{Z};\boldsymbol{{\alpha}}), \label{ipw}
\end{eqnarray}
where $g(\cdot)$ is some known function and $\boldsymbol{\hat{\alpha}}$ is the solution to some set of estimating equations that provides consistent estimators of the vector $\boldsymbol{\alpha}$ when the propensity score model is correctly specified, e.g. MLE score equations. If this model for the propensity score is correctly specified for confounding and we use it to obtain weights of the form $W(X,\boldsymbol{Z};\hat{\boldsymbol{\alpha}})$ as given above and additionally, we use the linear working outcome model as given in \eqref{linnoint}, then the coefficient for $X$ in the weighted OLS will be consistent for the ACE, i.e. $E\{Y(1)\} - E\{Y(0)\}$, regardless of the misspecification of the outcome model or the failure of the outcome model to contain all confounders. To be explicit, the estimator we are discussing is the $\beta$ within the set $\{\beta, \gamma_0, \boldsymbol{\gamma}\}$ that solves 
\begin{eqnarray*}
\frac{1}{n}\sum_{i=1}^{n}\left[
\begin{array}{c}W(X_i,\boldsymbol{Z}_i;\boldsymbol{\hat{\alpha}})\left\{\begin{array}{c} 1\\ X_i\\ m'(\boldsymbol{Z}_i, \boldsymbol{\gamma})\end{array}
\right \}[Y_i-\gamma_0-\beta X_i-m(\boldsymbol{Z}_i;\boldsymbol{\gamma})]
\end{array}\right]=0,
\end{eqnarray*}
where $m'(\boldsymbol{Z}, \boldsymbol{\gamma})=\frac{d m(\boldsymbol{Z}, \boldsymbol{\gamma})}{d\boldsymbol{\gamma}}$. The $\beta$ that is the solution to this set of estimating equations we call $\hat{\beta}^{ipw}$. Similarly, even if the working propensity score model is misspecified, if the outcome model is correctly specified for confounding, i.e if \eqref{eq:correct} and \eqref{eq:suff} hold, then $\hat{\beta}^{ipw}$ will be consistent for the ACE.  Proof of its double robustness is given in section S1 of the supporting information, but this is not a new result\cite{robinscomment2007}. 

The reason that $\hat{\beta}^{ipw}$ is consistent for the ACE without further regression standardization is because the average causal effect estimated via linear regression is collapsible. This will be an important consideration when we discuss generalized linear regression in the next section.  

Let us now consider a linear regression with interaction, 
\begin{eqnarray*}
E(Y|X,\boldsymbol{Z})=\gamma_0+\beta X+m(\boldsymbol{Z};\boldsymbol{\gamma}) + \gamma_{xz} X Z_1, \end{eqnarray*}
where $Z_1$ is one of the potential confounders we believe may have an interactive effect with the exposure for the outcome. The estimator we are considering uses the estimators of the vector $(\gamma_0, \beta, \boldsymbol{\gamma})$ that solve
\begin{eqnarray*}
\frac{1}{n}\sum_{i=1}^n \left[
\begin{array}{c}W(X_i,\boldsymbol{Z}_i;\boldsymbol{\hat{\alpha}})\left\{\begin{array}{c} 1\\ X_i\\ m'(\boldsymbol{Z}_i, \boldsymbol{\gamma})\\ X_iZ_{1i}\end{array}
\right \}[Y_i-\gamma_0-\beta X_i-m(\boldsymbol{Z}_i;\boldsymbol{\gamma})-\gamma_{xz} X_i Z_{1i}]
\end{array}\right]=0.
\end{eqnarray*}
We call the solution set  $\{\hat{\gamma}^{ipw*}_0, \hat{\beta}^{ipw*}, \boldsymbol{\hat{\gamma}}^{ipw*}\}$. In this case, the coefficient for $X$, $\hat{\beta}^{ipw*}$, is not a consistent estimator of the ACE even when both working models are correctly specified for confounding. In order to consistently estimate the ACE, we need to average over the interaction. One way of doing this is standardization. The standardization estimator, as outlined in Section \ref{not} above, of the expected value of $Y(x)$ is given by: 
 \begin{eqnarray*}
\hat{E}\{\hat{E}(Y|X=x,\boldsymbol{Z})\}=\frac{1}{n} \sum_{i=1}^n \hat{\gamma}_0^{ipw*}+ \hat{\beta}^{ipw*}x+m(x,\boldsymbol{Z}_i;\boldsymbol{\hat{\gamma}^{ipw*}}) + \hat{\gamma}_{xz}^{ipw*}x{Z}_{i1}.
\end{eqnarray*}
The contrast $\hat{E}\{\hat{E}(Y|X=1,\boldsymbol{Z})\} - \hat{E}\{\hat{E}(Y|X=0,\boldsymbol{Z})\}$ is a consistent estimator of ACE if either the outcome model or the propensity score model is correctly specified for confounding. A proof is given in section S1 of the supporting information. Note that, although the weights were used for the fitting of the coefficients, they should not be used in the standardization step. We provide example code to run this method in R in Section \ref{sec:example}  

\section{IPTW Canonical Link GLM}
Now consider the following generalized linear regression outcome model:
\begin{eqnarray*}
q\{E(Y|X,\boldsymbol{Z})\}=\gamma_0+\beta X+m(X,\boldsymbol{Z};\boldsymbol{\gamma}),
\end{eqnarray*}
where here $q$ is a canonical link and the arbitrary function $m(X,\boldsymbol{Z};\boldsymbol{\gamma})$ may or may not contain interactions between the exposure and the set of potential confounders, for example $\gamma_1Z_1+ \gamma_2XZ_1+\gamma_3Z_2$. We fit the GLM using IPTW MLE with estimated weights based on \eqref{ipw}, i.e. we used standard software to fit a GLM including weights, which weights each subject's data by $W(X,\boldsymbol{Z}, \boldsymbol{\hat{\alpha}})$. Then, if the outcome model is correctly specified for confounding or the weights are correctly specified for confounding, the standardization estimator is consistent for the ACE. A proof of this double robustness is given in section S1 of the supporting information.

Again the standardization estimator we are considering is based on estimators of the vector of $\{\gamma_0, \beta, \boldsymbol{\gamma}\}$ that solve 
\begin{eqnarray}
\frac{1}{n}\sum_{i=1}^n \left[\begin{array}{c}W(X_i,\boldsymbol{Z}_i;\boldsymbol{\hat{\alpha}})\left\{\begin{array}{c} 1\\ X\\ m'(X_i,\boldsymbol{Z}_i, \boldsymbol{\gamma})\\
\end{array}
\right \}[Y-q^{-1}\{\gamma_0+\beta X+m(X_i,\boldsymbol{Z}_i;\boldsymbol{\gamma})\}]
\end{array}\right]=0\label{glmcan}
\end{eqnarray}
where $m'(X_i,\boldsymbol{Z}_i, \boldsymbol{\gamma})=\frac{d m(X_i,\boldsymbol{Z}_i, \boldsymbol{\gamma})}{d\boldsymbol{\gamma}}$. We call the solution vector of equation \eqref{glmcan} $\{\hat{\gamma_0}^{ipw**}, \hat{\beta}^{ipw**}, \boldsymbol{\hat{\gamma}}^{ipw**}\}$. When we further standardize, our estimator of the ACE is given by $\hat{E}\{\hat{E}(Y|X=1, \boldsymbol{Z})\} -\hat{E}\{\hat{E}(Y|X=0, \boldsymbol{Z})\}$ where $$\hat{E}\{\hat{E}(Y|X=x, \boldsymbol{Z})\}=\frac{1}{n} \sum_{i=1}^n q^{-1}(\hat{\gamma}^{ipw**}_0+ \hat{\beta}^{ipw**}x+m(x,\boldsymbol{Z}_i;\boldsymbol{\hat{\gamma}}^{ipw**})).$$

The estimating equations in \eqref{glmcan} are the form of weighted score equations for an exponential family with fixed dispersion, as those used in GLM with a canonical link $q(\cdot)$. When the link is non-canonical, the resulting score equations may not be of this form, and thus there is no guarantee that misspecification of the outcome model, even when the weights model is correctly specified for confounding, will result in a consistent estimator for the ACE. Further explanation is provided in the supporting information section S3. This is not to say that DR estimators based on regression standardization cannot be constructed in such cases, simply that the estimators cannot be based on standardized IPTW noncanonical GLM. 

\subsection{Logistic regression example}
Consider a specific case of logistic regression, so the $q$ is the logit function and $Y$ is binary, where the mean model is
\begin{eqnarray*}
\mbox{logit}\{E(Y|X,\boldsymbol{Z})\}=\gamma_0+\beta X+m(x,\boldsymbol{Z};\boldsymbol{\gamma}),
\end{eqnarray*}
and we fit this model via weighted MLE, with weights based on the model in Equation \eqref{ipw}, giving us the coefficient vector $\{\hat{\gamma_0}^{ipw\dagger}, \hat{\beta}^{ipw\dagger}, \boldsymbol{\hat{\gamma}}^{ipw\dagger}\}$. When we further standardize, our estimator of the ACE is given by $\hat{E}\{\hat{E}(Y|X=1, \boldsymbol{Z})\} -\hat{E}\{\hat{E}(Y|X=0, \boldsymbol{Z})\}$ where $$\hat{E}\{\hat{E}(Y|X=x, \boldsymbol{Z})\}=\frac{1}{n} \sum_{i=1}^n q^{-1}(\hat{\gamma}^{ipw\dagger}_0+ \hat{\beta}^{ipw\dagger}x+m(x,\boldsymbol{Z}_i;\boldsymbol{\hat{\gamma}}^{ipw\dagger})).$$ This standardization estimator is consistent for the ACE if either the working outcome model or the working propensity score model is correctly specified for confounding. 

Thus, logistic regression, as with all noncollapsible link/estimand canonical link GLMs, is doubly robust after standardization, and the difference between the standardized probabilities is a consistent estimate of the ACE. Once again, we provide code for running logistic regression in combination with IPTW and using standardization to obtain the risk difference in Section \ref{sec:example}. It is of note here that, due to the noncollapsible link/estimand combination in logistic regression, the coefficients from the working outcome model will not be consistent estimates of any conditional causal odds ratio, unless the outcome model is correct specified for confounding, even if the propensity score is correct. These estimates will converge to the probability limit of the misspecified model in a randomized trial. Thus, with logistic regression, and all other noncollapsible link/estimand combination, standardization is required for doubly robust estimation of the ACE.

As we have demonstrated, when the ACE is the target of inference, one can use an IPTW canonical link GLM and regression standardization to obtain a DR estimator of the ACE.  Table \ref{alltab} provides a summary of the methods we have reviewed and if they are doubly robust for the ACE. 

\begin{table}[h]
    \centering
        \caption{Summary of the DR properties of weighting selected outcome regression methods \label{alltab}}
    \begin{tabular}{cccc}
   Working outcome model & Fitting Method & DR\\ 
         \hline
     linear regression no interaction& IPTW + OLS& Yes\\
       linear regression& IPTW + OLS + standardization& Yes\\
        canonical link GLM& IPTW + MLE + standardization&Yes\\
          \hline 
        non-canonical link GLM& IPTW + MLE& not generally\\
    non-canonical link GLM& IPTW + MLE + standardization& not generally\\
         \hline 
    \end{tabular}
\end{table}

\section{Inference}
We have so far only considered the consistency of the estimator and not how to perform inference. Several possible standard error estimators can be used; we review two of them here. The first is the whole procedure nonparametric bootstrap. This procedure refits the propensity score model and the outcome model after each bootstrap resample and then uses the 0.025 and 0.975 percentiles of the resulting estimates for the confidence interval. In simulations, we show that this procedure has proper coverage when the estimator is consistent, and we provide R code to run this procedure for the selected canonical link GLMs in combination with IPTW weights obtained from the logistic regression. 

The second is an influence function based standard error estimator that we have implemented and provide code to use in addition to the bootstrap. This estimator is derived and provided in Section S2 of the supporting information based on estimators given in \cite{tsiatis2006semiparametric}. It is of note that one could also use a sandwich based standard error formulation. However, if one were to account for the variation in all models in the formulation of the sandwich based standard error it would either be the same estimator as the influence function based standard error estimator, or be asymptotically equivalent.

The influence function based standard error estimator is constructed specifically to account for the uncertainty in incorrect working models. This can be seen in Table S1 of the supporting information which compares the empirical standard error over simulated replicates to estimators of the standard error. When both models are correctly specified for confounding, the extra terms in the asymptotic variance for estimation of the propensity score and outcome models approach zero, and hence the variance of estimator approaches that of the efficient influence function. In other words, when both models are correct, the estimator has the smallest possible variance, asymptotically, across all estimators that are asymptotically unbiased when the propensity score model is correctly specified. 

\section{Simulations}

\subsection{Data generation and analysis}

In all settings, for observations $i = 1, \ldots, n = 2000$ we generate independent covariates $Z_{i1}$ from a standard normal distribution and $Z_{i2}$ from a normal distribution with mean 1 and variance 1. Then the exposure $X_i$ is generated from a Bernoulli distribution with probability
\[
\mbox{expit}(-0.4 + 0.4 Z_{i1} + 0.28 Z_{i1}^2 + 0.4 Z_{i2}).
\]
The linear predictor for the outcome is generated as
\[
\eta_i = \gamma_0 + \beta X_i + \gamma_1 Z_{1i} + \gamma_2 Z_{i2}^2 + \gamma_3 Z_{i2},
\]
which is then used to generate the outcome $Y_i$ as a sample from a distribution $F$ with mean parameter $q^{-1}(\eta_i)$. We vary the distribution $F$ (which may have additional parameters not depending on the other covariates, as stated below) and link function $q$ as follows: 
\begin{enumerate}
    \item Gaussian: $F$ is normal with mean $q^{-1}(\eta_i)$, with variance 1, and $q(x) = x$ (identity). The coefficients $(\gamma_0, \beta, \gamma_1, \gamma_2, \gamma_3) = (-2, 2, 1, 0.4, 1.5)$. 
    \item Inverse Gaussian: $F$ is inverse Gaussian with mean $q^{-1}(\eta_i)$, shape parameter 2, and $q(x) = x^{-2}$.  The coefficients $(\gamma_0, \beta, \gamma_1, \gamma_2, \gamma_3) = (50, -200, 4, 10, 5)$.
    \item Poisson: $F$ is Poisson with mean $q^{-1}(\eta_i)$, and $q(x) = \log(x)$.  The coefficients $(\gamma_0, \beta, \gamma_1, \gamma_2, \gamma_3) = (0, 2, 0.1, 0.05, 0.4)$.
    \item Bernoulli: $F$ is Bernoulli with probability $q^{-1}(\eta_i)$, and $q(x) = \mbox{logit}(x)$.  The coefficients $(\gamma_0, \beta, \gamma_1, \gamma_2, \gamma_3) = (-2, 2, 1, 1, 4)$.
\end{enumerate}

Once data are generated, they are analyzed as follows: 

\begin{enumerate}
    \item A propensity score model is estimated using logistic regression with $X_i$ as the outcome and including predictors $Z_{i1}, Z_{i1}^2$ and $Z_{i2}$. This is the correctly specified propensity score model and it outputs predicted probabilities $\hat{p}^{(1)}_i$ of $X_i = 1$. 
    \item A propensity score model is estimated using logistic regression with $X_i$ as the outcome and including only $Z_{i1}$ as a predictor. This is the misspecified propensity score model and it outputs predicted probabilities $\hat{p}^{(2)}_i$ of $X_i = 1$. 
    \item For a given propensity score model that outputs $\hat{p}^{(j)}_i$, predicted probabilities of $X_i = 1$, we compute weights 
    \[
    W^{(j)}_i = \frac{X_i}{\hat{p}^{(j)}_i} + \frac{1 - X_i}{1 - \hat{p}^{(j)}_i}.
    \]
    \item A generalized linear outcome model with link function $q$ is estimated using maximum likelihood with weights $W^{(j)}_i$, with $Y_i$ as the outcome, and predictors $X_i, Z_{i1}, Z_{i1}^2$ and $Z_{i2}$. This is the correctly specified outcome model.
    \item A generalized linear outcome model with link function $d$ is estimated using iteratively reweighted least squares with weights $W^{(j)}_i$, with $Y_i$ as the outcome, and predictors $X_i, Z_{i1}$. This is the misspecified outcome model.
    \item For a given outcome model, regression standardization (unweighted) is used to estimate the ATE contrasting $X_i = 1$ to $X_i = 0$.
\end{enumerate}

We consider settings where $d$ is the identity, log, and logit link, and where the GLM is for Gaussian, Poisson, and binomial families, respectively. We also consider a binomial family with a log link to demonstrate what happens when the link is not canonical. For each setting, we do the analysis where the propensity score model is misspecified and the outcome model is correctly specified, where the propensity score model is correctly specified and the outcome model is misspecified, where both are misspecified, and where both are correctly specified. Our measure of interest is the average bias over 2000 replicates of the simulation: the mean of the estimated ATE minus the true ATE. We also report the standard deviation of the estimates over the simulation replicates. We also assess and compare the coverage of 95\% confidence intervals based on the standard error that does not take into account the estimation of the propensity scores, based on the quantiles of the bootstrap distribution, and using standard errors estimated using the influence function estimator.

We compared the above estimators to the doubly robust estimator as described in Funk et al.\cite{funk2011doubly}, but that was developed theoretically previously\cite{robins1994estimation}. In addition to a model for the propensity score that gives predicted probabilities $\hat{p}_i$, Funk et al.\cite{funk2011doubly} suggest to use two outcome models, one for the subgroup where $X_i = 0$ and another for the subgroup where $X_i = 1$, which yield predictions of the potential outcomes $\hat{Y}_{i0}$ and $\hat{Y}_{i1}$ for $i = 1, \ldots, n$. This is not required for the estimator to be doubly robust, it simply makes the modeling more flexible. This estimator for the ATE is given by: 
\[
\frac{1}{n}\sum_{i = 1}^n\frac{Y_i X_i}{\hat{p}_i} - \frac{\hat{Y}_{i1}(X_i - \hat{p}_i)}{\hat{p}_i} - \frac{Y_i (1 - X_i)}{1 - \hat{p}_i} - \frac{\hat{Y}_{i0}(X_i - \hat{p}_i)}{1 - \hat{p}_i}.  
\]
This estimator is also consistent if either the propensity score or outcome models are correctly specified. However, in practice, this estimator is often not implemented as it is described in the paper, see for example Karter et al.\cite{karter2021association} and Xie et al.\cite{xie2020comparative}, where instead of using an unweighted outcome model, the authors used a IPT weighted outcome model alone but cite Funk et al.\cite{funk2011doubly}. In selected settings, we compare these two estimators in terms of finite sample bias and efficiency. We show in Subsection S1.1 of the supporting information that the two estimators are asymptotically equivalent if both working models are correctly specified, and that in this case they are both efficient, having a variance given by the efficient influence function. 

We investigate one additional setting, which demonstrates that one cannot remove residual confounding by further adjusting an outcome model, when, for example, the working model for the weights is not correctly specified for confounding. For this example, we consider a binary confounder $V_i$ that is Bernoulli with probability 0.35, a continuous confounder $C_i$ that is standard normal, and the propensity model is 
\[
\mbox{expit}(-0.4 + 4 C_i + 0.28 C_i^2 + 0.4 V_i)
\]
and the outcome is generated as
\[
Y_i = -2 + 2 X_i + 1 C_i + 4.5 V_i. 
\]
Then we fit the propensity score model including only the predictor $V_i$ and the outcome model including only main effect terms for $X_i$ and $C_i$. In this case, both the propensity and outcome models are correctly specified statistically, but neither is correctly specified for confounding. 

R code and detailed documentation to reproduce the simulation study is available at \url{https://github.com/sachsmc/doubly-robust-GLM}

\subsection{Results}
\begin{longtable}{lrlrr}
\caption{Simulation Results from IPTW OLS and IPTW GLM. SD = standard deviation; coverage = confidence interval coverage percent; boot = nonparametric bootstrap; IF = influence function. \label{simresulttab1}} \\
\toprule
\multicolumn{5}{c}{Canonical link}\\
\midrule
type & true value & percent bias (SD)  & coverage boot & coverage IF \\ 
\midrule
\multicolumn{5}{l}{Generation: linear; analysis: ols weighted standardized} \\ 
\midrule
wrong outcome right weights & $2.00$ & 0.2 (0.06)& $94.7$ & 94.2\\ 
right outcome wrong weights & $2.00$ & 0.0 (0.05) & $95.5$ & 93.5\\ 
both wrong & $2.00$ & 37.5 (0.07)  & 0.0 & 0.0\\
right both & $2.00$ & 0.0 (0.05)  & 94.6 & 95.0 \\
\midrule
\multicolumn{5}{l}{Generation: log poisson; analysis: poisson weighted standardized} \\ 
\midrule
wrong outcome right weights & $10.94$ & 0.1 (0.17)  & $94.9$ & 94.7\\ 
right outcome wrong weights & $10.94$ & 0.1 (0.16)  & $94.6$& 94.3\\ 
both wrong & $10.94$ & 10.3 (0.21)  & $0.0$ & 0.0\\ 
right both & $10.94$ & 0.01 (0.16)  & $95.2$ & 94.7\\ 
\midrule
\multicolumn{5}{l}{Generation: logit binomial; analysis: logit binomial weighted standardized} \\ 
\midrule
wrong outcome right weights & $0.12$ & 0.0 (0.01)  & $95.1$ & 94.7\\ 
right outcome wrong weights & $0.12$ & -0.0 (0.01)  & $93.7$& 95.4\\ 
both wrong & $0.12$ & 96.3 (0.02)  & $0.0$ & 0.0\\ 
right both & $0.12$ & 0.0 (0.01)  & $95.2$ & 94.3\\ 
\midrule
\multicolumn{5}{l}{Generation: inverse gaussian; analysis: inverse gaussian weighted standardized} \\ 
\midrule
wrong outcome right weights & $-0.06$ & 0.0 (<0.01) & $94.5$ & 95.0\\ 
right outcome wrong weights & $-0.06$ & 0.0 (<0.01)  & $94.3$& 97.1\\ 
both wrong & $-0.06$ & 5.2 (<0.01) & $0.0$& 0.0\\ 
right both & $-0.06$ & 0.0 (<0.01)  & $94.8$& 95.1\\ 
\midrule
\multicolumn{5}{c}{Non-Canonical link}\\
\midrule
\multicolumn{5}{l}{Generation: logit binomial; analysis: log binomial weighted standardized} \\ 
\midrule
wrong outcome right weights & $0.33$ & -33.6 (0.03) & NA & NA\\ 
\bottomrule
\end{longtable}

\begin{table}[ht]
\caption{Standard deviations over the simulation replicates of the estimates \label{efffunk}}
\centering
\begin{tabular}{rrrrr}
  \hline
 & GLM linear & Funk linear & GLM logit & Funk logit \\ 
  \hline
  & \multicolumn{4}{c}{$n = 100$} \\
\hline
right outcome wrong weights & 0.224 & 0.224 & 0.059 & 0.059 \\ 
  both wrong & 0.381 & 0.381 & 0.077 & 0.077 \\ 
  wrong outcome right weights & 0.259 & 0.329 & 0.063 & 0.069 \\ 
\hline
  & \multicolumn{4}{c}{$n = 500$} \\
\hline
  right outcome wrong weights & 0.094 & 0.094 & 0.027 & 0.027 \\ 
  both wrong & 0.177 & 0.177 & 0.037 & 0.037 \\ 
  wrong outcome.right.weights & 0.120 & 0.141 & 0.030 & 0.030 \\ 
\hline
  & \multicolumn{4}{c}{$n = 1000$} \\
\hline
 right outcome wrong weights & 0.067 & 0.067 & 0.018 & 0.018 \\ 
  both wrong & 0.123 & 0.123 & 0.025 & 0.025 \\ 
  wrong outcome.right.weights & 0.081 & 0.089 & 0.020 & 0.020 \\ 
\hline
  & \multicolumn{4}{c}{$n = 2000$} \\
\hline
right outcome wrong weights & 0.047 & 0.048 & 0.013 & 0.013 \\ 
  both wrong & 0.085 & 0.085 & 0.018 & 0.018 \\ 
  wrong outcome right weights & 0.062 & 0.068 & 0.015 & 0.015 \\ 
  
   \hline
\end{tabular}
\end{table}

As can be seen in Table \ref{simresulttab1} the simulations support the theory that the IPTW Canonical link GLM and IPTW OLS estimators are doubly robust and that the whole procedure nonparametric bootstrap provides nominal coverage when the estimator is consistent, i.e. when one of the two working models is correctly specified for confounding. In the last section of Table \ref{simresulttab1} we demonstrate when the working outcome model is not using the canonical link, the result of the IPTW GLM estimator followed by standardization is not a doubly robust estimator. We also provide results showing the bias and coverage if both working models are misspecified. These results serve to demonstrate the worst case scenario where not only are both models misspecified but neither model even contains all of the confounders. In practice, if the measured confounders are sufficient for confounding control, the bias would be expected to be lower, although the estimator would still not be consistent. 

Table \ref{simresulttab1} also shows that the confidence intervals based on the bootstrap and the influence function based standard error both have the correct coverage when the estimator is consistent. The confidence intervals constructed using the naive standard error have severe under coverage. 

In the final scenario not included in Table \ref{simresulttab1}, where IPTW OLS was applied followed by standardization but both the propensity score model and the linear outcome model were correctly specified statistically but neither sufficiently for confounding control, we find an average bias (standard deviation) of 0.18 (0.12) and bootstrap 95\% confidence interval coverage of 70.5\%. In this scenario there were two confounders; the propensity score model was correctly specified for one confounder, while the outcome model was correctly specified for the other. This demonstrates that one cannot use the IPTW OLS, and through equivalence, the method of Funk et al.\cite{funk2011doubly}, to adjust for residual confounding, even if both models are correctly specified. In general, one working model must always be correctly specified for confounding for the estimator to be consistent. 

We also compare the finite sample efficiency of the method of Funk et al.\cite{funk2011doubly} and the IPTW GLM in Table \ref{efffunk}, and find even in finite samples the estimation procedures have very similar efficiency. This supports the use of whichever method is easier for the user, although, at least in our simulations, we do find slightly better finite sample performance of the IPTW GLM. 

\section{Real Data Example} \label{sec:example}
We illustrate the methods outlined in the simulation section with the low birth weight dataset, which is publicly available as part of the R package \texttt{AF} \citep{AF}. We use this example so that readers can use the code provided in the supplement to reproduce the results easily. The dataset includes information on 487 births from 189 mothers who have given birth to one or more children. As described by Hosmer and Lemeshow (2000)\citep{appliedlogistic}, the data are based on real data collected as part of a larger study at Baystate Medical Center in Springfield, Massachusetts. The dataset includes information on the birth order (i.e., first born, second born, etc.), whether or not the mother is a smoker, the race of the mother with levels white, black, or other, the age of the mother in years, the weight of the mother in pounds at last menstrual period, and the birth weight of the newborn in grams.

First, we analyze the data using a linear model where smoking is the exposure, and birth weight in grams is the outcome. The potential confounders are race (white, black, other), age of the mother in years, and weight of the mother at last menstrual period (in pounds). The working outcome model includes an intercept, an indicator of smoking, an indicator of black race, and an indicator of other race, age as a linear term as well as age squared, weight as a linear term as well as weight squared, interactions between the smoking indicator and the race indicators, and interactions terms between smoking and age and  smoking and weight. The working propensity score model is a logistic regression for the outcome of smoking, including an intercept, the indicators of race black, and race other, a linear term for age and age squared, a linear term for weight and weight squared, interactions between the race indicators and age and weight, an interaction between age and weight, and the three-way interaction between the race indicators, age, and weight. The code to implement OLS weight with standardization and estimator described by Funk et al.\cite{funk2011doubly} using these models is provided in the supporting materials. We use the \texttt{glm} function in base R specifying a binomial family with a logit link to fit the propensity score model. We then use the \texttt{predict} function from the \texttt{stats} R package using option \texttt{type="response"} to obtain predicted probabilities for each subject. We then constructed the IPTW as outlined above, weighting those observed to have the exposure by the inverse probability of being exposed and those that are observed to not be exposed by the inverse probability of not being exposed. We include these weights, using the weights option in the \texttt{GLM} function in base R with family Gaussian and the identity link for the outcome model. We then again use the \texttt{predict} function using the resulting model object, but on two new data sets, which are the same as the original data set but setting the exposure to 1 or 0 for all subjects, respectively. Note that the IPTW are not used during this procedure. Having obtained these two sets of predictions we take the average over the subjects and take the difference between the all exposed estimate and the all unexposed estimate to obtain our estimate of the ATE. We note that if one uses the \texttt{stdReg} R package, as we have in the simulations, one must set the \texttt{prior.weights} element of the GLM object to a vector of 1s before passing it to the standardization function in order to get estimates for the observed and not the hypothetically randomized population. Finally, to obtain 95\% confidence intervals we either implement the influence function based standard error estimate or we bootstrap the full procedure as outlined in the inference section above. 

Using this procedure, the IPTW OLS followed by standardization, gives an estimated ATE of $-$225 grams, with an influence function-based 95\% confidence interval of ($-$446 to $-$1.2). We note that if one does not include interactions between the exposure and potential confounders, then the coefficient estimate for the exposure is a doubly robust estimate of the ATE. We also bootstrap the entire procedure, including the estimation of the propensity score model, to estimate confidence intervals; this gives a bootstrap 95\% confidence interval of ($-$387 to $-$46) grams. Thus, regardless of the standard error estimator used, the inference suggests that smoking is associated with a significant reduction in birth weight. We compare the IPTW OLS standardization estimate to the estimate given by using the method of \citep{funk2011doubly}, this is $-$226 grams with a bootstrap 95\% CI of ($-$390 to $-$52). Although the confidence intervals are somewhat different, all inference agrees that we can reject the null of no effect of smoking on birth weight, with both the IPTW OLS and the method of \citep{funk2011doubly} providing similar estimates for the ATE. 

We also analyze the binary outcome of an indicator of low birth weight (less than 2500 grams) using a similar procedure. Here, we could continue to use an IPTW OLS, either followed by standardization if there are interactions or the coefficient estimate for the exposure directly if there is no interaction. However, as it is unlikely that this can be correctly specified for confounding over the full range of continuous covariates, like age or weight, we use the binomial family and the logit canonical link. The ATE in this analysis is thus the difference in probabilities of low birth weight, comparing smokers to nonsmokers. The propensity score model remains as outlined above, and the outcome model, although now for a binary outcome and using a logit link, includes all of the same variables via the same interactions and functional forms as outlined in the linear setting. We use the same fitting procedure as outlined above, simply replacing the Gaussian GLM outcome model with the logistic GLM outcome model. 

Using the IPTW logistic GLM followed by standardization, we get an estimate of 0.13 for the risk difference with an influence function-based 95\% confidence interval of (0.01 to 0.27) and a bootstrap 95\% confidence interval of (0.03 to 0.24). Using the method of \citep{funk2011doubly} we get an estimate of 0.14 with a bootstrap 95\% confidence interval of (0.03 to 0.24). Again the inference remains the same for all methods, and suggests a significant effect of smoking on the risk of low birth weight.

\section{Discussion}
Although it is not a new result, we have demonstrated via simulation and analytical proof that a canonical link GLM fit via maximum likelihood, where one directly weights the score equations, followed by regression standardization delivers a doubly robust estimator for the ACE. Although we discussed it separately, this result also includes IPTW OLS and IPTW OLS followed by regression standardization. We have demonstrated via simulation that, when using this type of DR estimator, one of the working models must be correctly specified for confounding. This is true even if both models are correctly specified and contain between them all confounders, as in the example in the simulations. Thus, one cannot adjust for residual confounding not captured in the propensity score, for example, by adjusting only for that residual confounding in the outcome model. Additionally, although we have focused throughout on causal estimands, the same reasoning for double robustness can be applied to a marginal but adjusted association, which may be a more feasible target when unmeasured confounders are present. 

We have shown that the IPTW GLM and IPTW OLS, in addition to regression standardization, are asymptotically equivalent to the method outlined in Funk et al.\cite{funk2011doubly} and found no notable differences in simulations in the resulting estimates or the efficiency in practice in finite samples. Although both methods are DR, it may sometimes be easier to use the IPTW GLM method rather than constructing the estimator given in Funk et al.\cite{funk2011doubly}. It has the further advantage of guaranteeing ACE estimates in the parameter space (e.g., $[-1,1]$ for a dichotomous outcome).'

We have shown that bootstrap will provide proper inference for the ACE using IPTW GLM or OLS followed by standardization if the estimator is consistent for the ACE, i.e. if one of the two working models is correctly specified for confounding. We have also derived and implemented a closed-form standard error estimator based on the influence function, which we have also demonstrated in simulations to provide nominal coverage in finite samples. The code for this estimator is provided in the supporting information.

We have provided an example code for using the IPTW GLM and the method outlined in Funk et al.\cite{funk2011doubly} to obtain a DR estimator of the ACE and the two inference methods. For applied researchers who wish to use a DR estimator, but do not want to or are uncomfortable using a method they are unfamiliar with, for example, TMLE or double machine learning, IPTW GLM may be a useful tool for analysis. We note that, although we provide code for these estimators, we do not encourage their use above estimators, such as TMLE that provide a higher likelihood of getting a correctly specified outcome or propensity model by supporting more flexible modeling. However, we also note that the more flexible versions of TMLE are generally more computationally and conceptually complex. As a result, this approach can have higher sample size requirements and does not automatically outperform the more parametric approaches. We have also not considered more complex data structures, such as missing or longitudinal data. In these settings, once again, other methods that have been designed to handle such structures are likely better approaches than ad hoc modifications of IPTW GLM.  

Finally, we caution the reader that, although IPTW canonical link GLM and IPTW OLS followed by standardization delivers doubly robust estimators when the GLM is fitted via weighted MLE, not all methods of estimation that combine a working outcome model and a working propensity score model result in a doubly robust estimator. Although it is beyond the scope of this paper, this is particularly true for propensity score weighting of common survival models. This is well demonstrated by the IPTW non-canonical link GLM estimator, which is not doubly robust, even though it uses maximum likelihood for fitting the model, the subtle difference in the resulting score estimating equations eliminates the ability of the weights to compensate for a misspecified outcome model.  

\section*{ACKNOWLEDGMENTS}
EEG and MCS were supported in part by Novo Nordisk Fonden Grant  NNF22OC0076595.

\section*{Data Availability}
The low birth weight dataset used in the real data example is publicly available as part of the R package \texttt{AF} \citep{AF}.

\bibliographystyle{plainnat}
\bibliography{refs}
\end{document}